\documentclass[twocolumn,secnumarabic,amssymb, nobibnotes, showpacs, aps, prc,superscriptaddress, 10pt]{revtex4-1}

\usepackage{graphicx}
\usepackage{longtable}
\usepackage{lineno}
\begin{document}
%\linenumbers
\title{Light charged clusters emitted in 32 MeV/nucleon $^{136,124}$Xe+$^{124,112}$Sn reactions: 
chemical equilibrium, $^{3}$He and $^{6}$He production.}
\author{R. Bougault}
\affiliation{Normandie Univ, ENSICAEN, UNICAEN, CNRS/IN2P3, LPC Caen, F-14000 Caen, France}
\author{E. Bonnet}
\affiliation{SUBATECH UMR 6457, IMT Atlantique, Universit\'e de Nantes, CNRS-IN2P3, 44300 Nantes, France}
\author{B. Borderie}
\affiliation{Institut de Physique Nucl\'eaire, CNRS/IN2P3, Univ. Paris-Sud, Universit\'e Paris-Saclay, F-91406 Orsay cedex, France}
\author{A. Chbihi}
\affiliation{Grand Acc\'el\'erateur National d'Ions Lourds (GANIL), CEA/DRF–CNRS/IN2P3, Bvd. Henri Becquerel, 14076 Caen, France}
\author{D. Dell'Aquila}
\affiliation{Institut de Physique Nucl\'eaire, CNRS/IN2P3, Univ. Paris-Sud, Universit\'e Paris-Saclay, F-91406 Orsay cedex, France}
\affiliation{Dipartimento di Fisica 'E. Pancini' and Sezione INFN, Universit\'a di Napoli 'Federico II', I-80126 Napoli, Italy}
\author{Q. Fable}
\affiliation{Grand Acc\'el\'erateur National d'Ions Lourds (GANIL), CEA/DRF–CNRS/IN2P3, Bvd. Henri Becquerel, 14076 Caen, France}
\author{L. Francalanza}
\affiliation{Dipartimento di Fisica 'E. Pancini' and Sezione INFN, Università di Napoli 'Federico II', I-80126 Napoli, Italy}
\author{J.D. Frankland}
\affiliation{Grand Acc\'el\'erateur National d'Ions Lourds (GANIL), CEA/DRF–CNRS/IN2P3, Bvd. Henri Becquerel, 14076 Caen, France}
\author{E. Galichet}
\affiliation{Institut de Physique Nucl\'eaire, CNRS/IN2P3, Univ. Paris-Sud, Universit\'e Paris-Saclay, F-91406 Orsay cedex, France}
\affiliation{Conservatoire National des Arts et Metiers, F-75141 Paris Cedex 03, France}
\author{D. Gruyer}
\affiliation{Dipartimento di Fisica, Universit\'a di Firenze, via G. Sansone 1, I-50019 Sesto Fiorentino (FI), Italy}
\author{D. Guinet}
\affiliation{IPNL/IN2P3 et Universit\'e de Lyon/Universit\'e Claude Bernard Lyon1, 43 Bd du 11 novembre 1918 F69622 Villeurbanne Cedex, France}
\author{M. Henri}
\affiliation{Normandie Univ, ENSICAEN, UNICAEN, CNRS/IN2P3, LPC Caen, F-14000 Caen, France}
\author{M. La Commara}
\affiliation{Dipartimento di Fisica 'E. Pancini' and Sezione INFN, Università di Napoli 'Federico II', I-80126 Napoli, Italy}
\author{N. Le Neindre}
\affiliation{Normandie Univ, ENSICAEN, UNICAEN, CNRS/IN2P3, LPC Caen, F-14000 Caen, France}
\author{I. Lombardo}
\affiliation{Dipartimento di Fisica 'E. Pancini' and Sezione INFN, Università di Napoli 'Federico II', I-80126 Napoli, Italy}
\author{O. Lopez}
\affiliation{Normandie Univ, ENSICAEN, UNICAEN, CNRS/IN2P3, LPC Caen, F-14000 Caen, France}
\author{L. Manduci}
\affiliation{Ecole des Applications Militaires de l'Energie Atomique, BP 19 50115, Cherbourg Arm\'ees, France}
\affiliation{Normandie Univ, ENSICAEN, UNICAEN, CNRS/IN2P3, LPC Caen, F-14000 Caen, France}
\author{P. Marini}
\affiliation{CEA, DAM, DIF, F-91297 Arpajon, France}
\author{M. P\^arlog}
\affiliation{Normandie Univ, ENSICAEN, UNICAEN, CNRS/IN2P3, LPC Caen, F-14000 Caen, France}
\author{R. Roy}
\affiliation{Laboratoire de Physique Nucl\'eaire, Universit\'e Laval, Qu\'ebec, Canada G1K 7P4}
\author{P. Saint-Onge}
\affiliation{Laboratoire de Physique Nucl\'eaire, Universit\'e Laval, Qu\'ebec, Canada G1K 7P4}
\affiliation{Grand Acc\'el\'erateur National d'Ions Lourds (GANIL), CEA/DRF–CNRS/IN2P3, Bvd. Henri Becquerel, 14076 Caen, France}
\author{G. Verde}
\affiliation{Institut de Physique Nucl\'eaire, CNRS/IN2P3, Univ. Paris-Sud, Universit\'e Paris-Saclay, F-91406 Orsay cedex, France}
\affiliation{INFN - Sezione Catania, via Santa Sofia 64, 95123 Catania, Italy}
\author{E. Vient}
\affiliation{Normandie Univ, ENSICAEN, UNICAEN, CNRS/IN2P3, LPC Caen, F-14000 Caen, France}
\author{M. Vigilante}
\affiliation{Dipartimento di Fisica 'E. Pancini' and Sezione INFN, Università di Napoli 'Federico II', I-80126 Napoli, Italy}
%\date{May}\normalfont\Large\LARGE\normalfont
\vspace{0.5cm}
\collaboration{INDRA collaboration}
\noaffiliation
\begin{abstract}
Nuclear particle production from peripheral to central events is presented. N/Z gradient between projectile and target is studied using the fact that two reactions have the same projectile+target N/Z and so the same neutron to proton ratio for the combined system.
Inclusive data study in the forward part of the center of mass indicates that N/Z equilibration between the projectile-like and the target-like is achieved for central collisions. Particles are also produced from mid-rapidity region.  
$^{3}$He mean pre-equilibrium character is evidenced and $^{6}$He production at mid-rapidity implies a neutron enrichment
phenomenon of the projectile target interacting zone.
\end{abstract}
\date{March 8, 2017}
\maketitle
%%%%%%%%%%%%%%%%%%%%%%%%%%%%%%%%%%%%%%%%%%%%%%%%%%%%%%%%
%                                                                                        INTRODUCTION
%%%%%%%%%%%%%%%%%%%%%%%%%%%%%%%%%%%%%%%%%%%%%%%%%%%%%%%%
\section{Introduction}
The motivation for colliding nuclei at sufficient energy is to understand transport properties
of nuclear matter and also study nuclear matter under extreme conditions \cite{EPJA30All}. These are the two
inseparable faces of Heavy-Ion reaction research which are related to Dynamical and Statistical Physics.
\par
The knowledge concerning the achieved degree of equilibration between the two main collision partners 
is connected to both aspects and many degrees of freedom are concerned with equilibration.
At low bombarding energy \cite{GalinZPhys278} it has been shown that the N/Z ratio, isospin, is the fastest to be equilibrated. 
At higher bombarding energy, few tens MeV/nucleon, chemical equilibrium is driven by isospin diffusion in presence of an isospin gradient between the projectile and the target \cite{LiuPRC76} and by isospin drift sparked by a density gradient which occurs when a low density contact zone is created between the two partners \cite{BaranPR410}. 
The degree of chemical equilibration which is the N/Z balance between the projectile and the target is thus correlated to the 
interaction time between the two reaction partners, i.e the time left to isospin drift and diffusion mechanisms to be fully efficient.
In recent years many studies have been published concerning isospin transport in heavy ion reactions because it is connected to the knowledge of the isospin part of the equation of state and essential to resolve several issues in astrophysics (see \cite{EPJA50All} and references therein).
\par
Our experimental analysis concerning isospin equilibrium may be compared with results obtained with symmetric systems at comparable bombarding energy \cite{KeksisPRC81} \cite{SunPRC82}. We have here followed a different approach which is based 
on direct and simple observations: the presented analysis does not contain any data selection and presents mean value behaviors against reaction centrality.     
%%%%%%%%%%%%%%%%%%%%%%%%%%%%%%%%%%%%%%%%%%%%%%%%%%%%%%%
%%%%%%%%%%%%                                                   EXPERIMENTAL DETAILS
%%%%%%%%%%%%%%%%%%%%%%%%%%%%%%%%%%%%%%%%%%%%%%%%%%%%%%%%
\section{Experimental details}
The $4\pi$ multi-detector INDRA 
\cite{Pou95} was used to study four reactions with beams of $^{136}$Xe and $^{124}$Xe, 
accelerated at 32 MeV/nucleon, and thin (530 $\mu$g/cm$^{2}$) targets of $^{124}$Sn and $^{112}$Sn. 
Recorded event functionality was activated under a triggering factor based on a minimum number of fired 
detectors (M$_{trigger}^{min}$) over the detector acceptance (90\% of $4\pi$). 
During the experiment, performed at GANIL (Caen, France), 
minimum bias (M$_{trigger}^{min}$=1) and exclusive (M$_{trigger}^{min}$=4) data were recorded.\par
For light charged particle identification, detailed in \cite{Pou95}, two types
of thresholds for H and He elements are used in this study: (i) for individual isotope characteristics, as multiplicities, only fully identified particles (A and Z) are taken into account, 
(ii) for global light charged particle variables, as light charged particle total transverse energy, 
solely Z identified particles are also included. For solely Z identified particles,
A=1, 4 is respectively assigned to all H, He elements and this increases by about 4\%, 3\% respectively the 
studied $^{1}$H,  $^{4}$He populations as compared to fully identified one.\par 
This study is limited to the forward part of the center of mass (hereinafter called c.m.) and all figures, tables and measured quantities 
are related to this half hemisphere. It is thus focused on the evolution of projectile-like fragment isotopic content for which
the INDRA multi-detector, for these reactions, possesses excellent detection performances. The two studied
reactions
($^{124}$Xe+$^{124}$Sn and  $^{136}$Xe+$^{112}$Sn) 
were chosen to study the path towards chemical equilibrium
since their projectile+target combined systems are identical.
%%%%%%%%%%%%%%%%%%%%%%%%%%%%%%%%%%%%%%%%%%%%%%%%%%%%%%%%
%%%%%%%%%%%%                                          DETECTED CROSS SECTIONS
%%%%%%%%%%%%%%%%%%%%%%%%%%%%%%%%%%%%%%%%%%%%%%%%%%%%%%%%
\section{Detected cross-sections}
For minimum bias trigger recorded events, the detected reaction cross-section is given in table \ref{TableCrossSection}. The corresponding maximum impact parameter is also indicated for each 
studied reaction.
%%%%%%%%%%%
\begin{table}[h]
\centering
\begin{tabular}{|c|c|c|c|}
\hline	
\raisebox{-0.5ex}{} & {Detected} & {b$_{max}$} & {Detected}\\
\raisebox{-0.5ex}{} & {cross-section} & {} & {cross-section}\\
\raisebox{-0.5ex}{} & {(mb)} & {(fm)} & {(\%)}\\
\hline	  
\raisebox{-0.5ex}{$^{124}$Xe+$^{112}$Sn} & {3550 mb} &	{10.6 fm} & {64\%}\\
\hline	  
\raisebox{-0.5ex}{$^{124}$Xe+$^{124}$Sn} & {3870 mb} &	{11.1 fm} & {67\%}\\
\hline
\raisebox{-0.5ex}{$^{136}$Xe+$^{112}$Sn} & {4145 mb} &	{11.5 fm} & {72\%}\\
\hline
\raisebox{-0.5ex}{$^{136}$Xe+$^{124}$Sn} & {4500 mb} &	{12 fm} & {74\%}\\		  	  
\hline		  	    
\end{tabular}
\caption{Detected reaction cross-section and corresponding maximum impact parameter for each studied reaction. 
Last column corresponds to detected cross-section values relative to predictions of \cite{KoxCrossSection}. The minimum bias samples were used.}
\label{TableCrossSection}
\end{table}
%%%%%%%%%%%
Events with no detected light charged particle (H and He isotopes, hereinafter called lcp) in the forward part of the 
center of mass were eliminated. This condition ensures elimination of elastic scattering process but excludes also very 
peripheral reactions which lead to solely neutron evaporation. Uncharged particles are not detected by the apparatus 
nevertheless measured values indicate that a large fraction of the reaction cross-section has been recorded when 
compared to predictions of \cite{KoxCrossSection} (see table \ref{TableCrossSection}).
N-rich systems are slightly better detected than n-poor systems. This is probably caused by a smaller effect of the 
beam pipe dead zone for n-rich systems since very peripheral event neutron emission tends to deflect excited projectile prior lcp evaporation.  
\par  
%%%%%%%%%
%\begingroup
%\squeezetable
\begin{table}[h]
\centering
\begin{tabular}{|c|c|c|}
\hline		  
{} & \small{124+112} & \small{124+124}\\ 
\hline	    
\raisebox{-0.5ex}{$^{1}$H} & 
\raisebox{-0.5ex}{7963 ($\pm$7) mb} & \raisebox{-0.5ex}{7167 ($\pm$5) mb}\\ 
\hline
\raisebox{-0.5ex}{$^{2}$H} & 
\raisebox{-0.5ex}{2485 ($\pm$4) mb} & \raisebox{-0.5ex}{2714 ($\pm$3) mb}\\ 
\hline
\raisebox{-0.5ex}{$^{3}$H} & 
\raisebox{-0.5ex}{1342 ($\pm$3) mb} & \raisebox{-0.5ex}{1780 ($\pm$2) mb}\\ 
\hline		  
\raisebox{-0.5ex}{$^{3}$He} &  
\raisebox{-0.5ex}{572 ($\pm$2) mb} & \raisebox{-0.5ex}{491 ($\pm$1) mb}\\
\hline		  
\raisebox{-0.5ex}{$^{4}$He} &  
\raisebox{-0.5ex}{6992 ($\pm$6) mb} & \raisebox{-0.5ex}{7257 ($\pm$5) mb}\\
\hline		  
\raisebox{-0.5ex}{$^{6}$He} &  
\raisebox{-0.5ex}{109 ($\pm$1) mb} & \raisebox{-0.5ex}{147 ($\pm$1) mb}\\
\hline
\raisebox{-0.5ex}{Total} &  \raisebox{-0.5ex}{19463 ($\pm$23) mb} & \raisebox{-0.5ex}{19556 ($\pm$17) mb}\\
\hline
\hline		  
{} & \small{136+112} & \small{136+124}\\ 
\hline	    
\raisebox{-0.5ex}{$^{1}$H} & 
\raisebox{-0.5ex}{6621 ($\pm$7) mb} & \raisebox{-0.5ex}{6241 ($\pm$5) mb}\\
\hline
\raisebox{-0.5ex}{$^{2}$H} &  
\raisebox{-0.5ex}{2773 ($\pm$5) mb} & \raisebox{-0.5ex}{3092 ($\pm$4) mb}\\
\hline
\raisebox{-0.5ex}{$^{3}$H} & 
\raisebox{-0.5ex}{1965 ($\pm$4) mb} & \raisebox{-0.5ex}{2612 ($\pm$3) mb}\\
\hline		  
\raisebox{-0.5ex}{$^{3}$He} &  
\raisebox{-0.5ex}{416 ($\pm$2) mb} & \raisebox{-0.5ex}{397 ($\pm$1) mb}\\
\hline		  
\raisebox{-0.5ex}{$^{4}$He} &  
\raisebox{-0.5ex}{7005 ($\pm$8) mb} & \raisebox{-0.5ex}{7501 ($\pm$6) mb}\\
\hline		  
\raisebox{-0.5ex}{$^{6}$He} &  
\raisebox{-0.5ex}{163 ($\pm$1) mb} & \raisebox{-0.5ex}{239 ($\pm$1) mb}\\
\hline
\raisebox{-0.5ex}{Total} & \raisebox{-0.5ex}{18943 ($\pm$27) mb} & \raisebox{-0.5ex}{20083 ($\pm$20) mb}\\
\hline		  		  	    
\end{tabular}
\caption{Forward c.m. detected lcp cross-sections for $(\Sigma E_{t})_{FwCM}^{lcp}>0$. Values are given for each studied reaction (124+112 refers to $^{124}$Xe+$^{112}$Sn, etc...) and each type of fully identified lcp. Last line corresponds to the sum of lcp cross-sections for each system. Statistical errors are given in parenthesis. The minimum bias samples were used.}
\label{TableCrossSectionlcp}
\end{table}
%\endgroup
%%%%%%%%%%%%%%%
Forward c.m. detected cross-sections are given in table \ref{TableCrossSectionlcp} for each type of lcp. The total lcp cross-section values are presented in the last line for each system and they show an almost, within 6\%, system independence while the chemistry of lcp production is largely system dependent. 
As a matter of fact, increasing the neutron 
richness of the combined projectile-target system, we notice the following points:
(i) n-poor particle ($^{1}$H, $^{3}$He) production decreases,
(ii) $^{4}$He production is not strongly sensitive to N/Z change and it becomes the most produced particle,
(iii) $^{2}$H production increases while almost an equal value is measured for the two identical combined-N/Z systems,
(iv) n-rich particle ($^{3}$H, $^{6}$He) production largely increases - the cross-section values are doubled. 
Thus changing projectile and target N/Z, isotope production cannot be summed up in solely neutron production differences and
these chemical modifications are the basic scope of this study. 
%%%%%%%%%%%%%%%%%%%%%%%%%%%%%%%%%%%%%%%%%%%%%%%%%%%%%%%%
%%%%%%%%%%%%               IMPACT PARAMETER EVALUATOR AND MIXED SAMPLES
%%%%%%%%%%%%%%%%%%%%%%%%%%%%%%%%%%%%%%%%%%%%%%%%%%%%%%%%
\section{Impact parameter evaluator and mixed samples}
\begin{figure}[ht]
\centering
\resizebox{0.45\textwidth}{!}{%
   \includegraphics{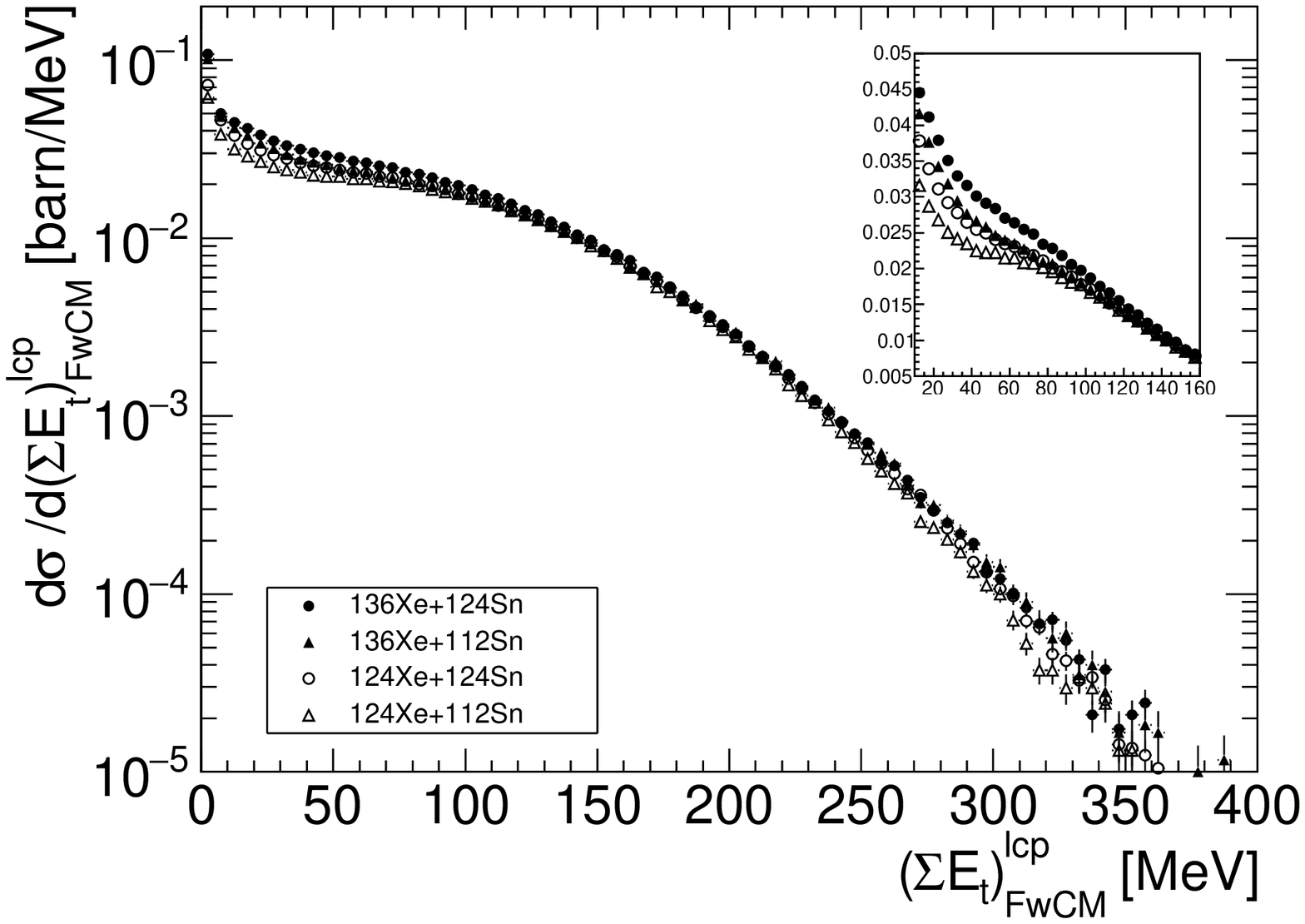}
   % code figure FiguresArbres_XeSnRunsSingles.C pour root : 1 
   } 
\resizebox{0.45\textwidth}{!}{%
   \includegraphics{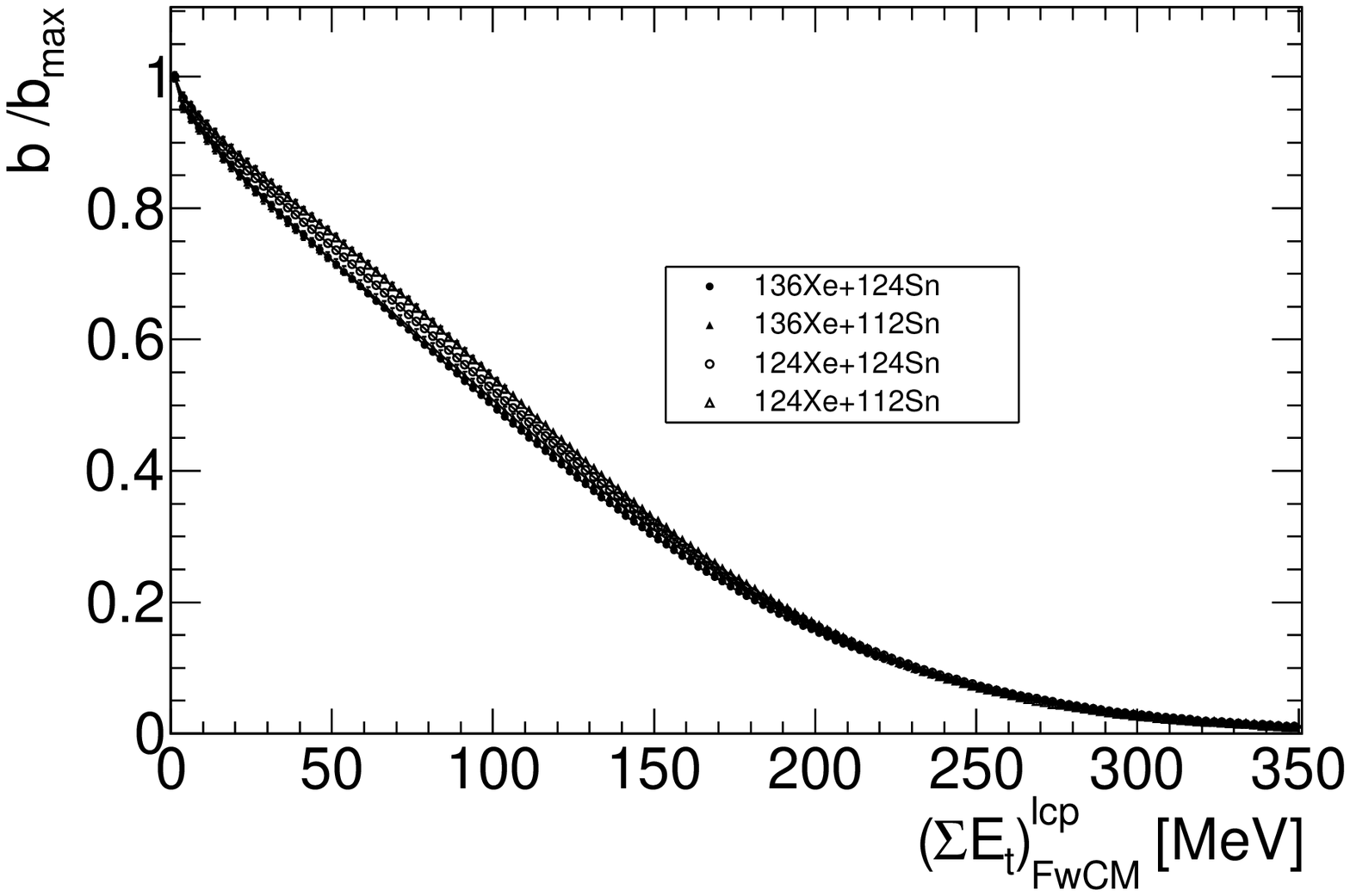}
   % code figure FiguresArbres_XeSnRunsSingles.C pour root : 1.1 
   }   
   \caption{Top: sum of light charged particle transverse energy distribution for the four studied reactions.  
Light charged particles refer to H and He isotopes detected in the forward part of the center of mass 
(minimum bias sample). 
A zoom of the left part of the figure is also shown. Bottpm: relationship between the impact parameter 
evaluator and the reduced impact parameter 
for the four studied reactions (minimum bias samples).}
   \label{EtlcpAvecZoom_bsurbmax}
\end{figure}

Light charged particle total transverse energy ($(\Sigma E_{t})_{FwCM}^{lcp}$) distributions are displayed in 
figure \ref{EtlcpAvecZoom_bsurbmax}-left
for the four reactions under minimum bias triggering condition.
%%%%%%%%%%%
%%%%%%%%%%%
For very low $(\Sigma E_{t})_{FwCM}^{lcp}$ values, the cross-section is system dependent, it increases with 
the mass of the system (see inset panel). For a transverse energy greater than 60 MeV the $^{136}$Xe+$^{112}$Sn and 
$^{124}$Xe+$^{124}$Sn distributions are identical. $^{124}$Xe+$^{112}$Sn distribution merges the 
two identical ones at 100 MeV and finally all distributions behaves the same for values greater than 150 MeV. 
This reflects the behavior of particle production related to nucleon exchanges between the projectile and the target 
as a function of impact parameter and the $(\Sigma E_{t})_{FwCM}^{lcp}$ observable has been used, in the following, 
as an impact parameter evaluator \cite{MetivierNPA672}.\par 
The relationship between $(\Sigma E_{t})_{FwCM}^{lcp}$ and the reduced impact parameter is given in 
figure \ref{EtlcpAvecZoom_bsurbmax}-right 
using the technique of reference \cite{CavataPRC42} with minimum bias trigger condition data. 
It is observed that the 150 MeV value of the lcp transverse energy from which all $(\Sigma E_{t})_{FwCM}^{lcp}$ 
distributions are identical corresponds to a 0.3 reduced impact parameter value, thus central collision events.
\par
\begin{figure}[htbp]
\centering
%\resizebox{0.98\textwidth}{!}{%
\resizebox{0.45\textwidth}{!}{%
   \includegraphics{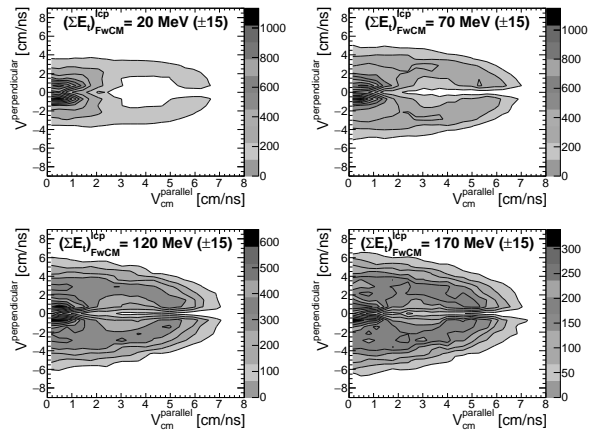}
   % code figure FiguresArbres_XeSnRunsSingles.C pour root : 4.165 
   }
   \caption{ $^{2}$H Galilean invariant center of mass velocity contour plots for different impact parameter evaluator ranges (z-scale: counts per cm$^{2}$/ns$^{2}$, bin size: 0.32x0.72 cm$^{2}$/ns$^{2}$). The $^{124}$Xe+$^{124}$Sn minimum bias sample was used.}
   \label{VparVperEt}    
\end{figure}

%\section{Lcp production mechanism}
Figure \ref{VparVperEt} presents $^{2}$H Galilean invariant velocity plots for different impact parameter evaluator gates from peripheral to central $^{124}$Xe+$^{124}$Sn collisions. 
%%%%%%%%%%%
%%%%%%%%%%%
Emission from excited projectile-like fragment (PLF) is observed and is centered on a velocity value which evolves with impact parameter. This emission is increasing with the damping of the PLF velocity indicating an increase of the PLF excitation energy. The particle production whose velocity is located between the Projectile-like and the target-like fragment (TLF) velocities is referred as mid-rapidity production and it reflects the dynamical nature of the process which occurs between the two partners during the collision \cite{EPJA30DiToro}. At this bombarding energy the mid-rapidity region is populated by PLF and TLF emission and remnants of the contact region between the PLF and TLF where exchange of nucleons occurs between the two main partners. Mid-rapidity, or neck, has been evidenced in heavy-ion collisions (\cite{EPJA30DiToro} and references therein)
and was studied for Xe+Sn system at different bombarding energies \cite{PlagnolPRC61}. Projectile/target nucleon exchange and mid-rapidity zone chemical composition is mainly governed by diffusion and drift isospin transport phenomena at this bombarding energy \cite{LiuPRC76} \cite{BaranPR410}. 
\par
Figure \ref{VparVperEt} also indicates that focusing on lcp production at different center of mass polar angle ranges it is possible to approximately select projectile-like de-excitation (0$^{\rm{o}}$-30$^{\rm{o}}$) or mid-rapidity (60$^{\rm{o}}$-90$^{\rm{o}}$) populations. 
Whatever the lcp production mode (evaporation, simultaneous production, secondary decay,...) the global production is linked to the (neutron, proton) composition of the concerned angular zone and
these angular selections will be used in the following for average behavior of both mentioned populations.
More sophisticated selection methods exist for projectile-like de-excitation characterization but they only
apply to very exclusive data \cite{Vient2016}.\par
%%%%%%%%%%%%%%%%%%%%%%%%%%%%%%%%%%%%%%%%%%%%%%%%%
%%                                      Minimum biais and mixed samples
%%%%%%%%%%%%%%%%%%%%%%%%%%%%%%%%%%%%%%%%%%%%%%%%%
%\section{Minimum biais and mixed samples}
A $4\pi$ multi-detector is necessary to study reaction mechanisms but it implies a large flux of data during the experiment. 
This large flux generally causes large acquisition dead time. This issue is solved by reducing the beam intensity but longer 
run periods are necessary. A good compromise is to use exclusive data recording but one has to be able to verify that 
a correct sampling is achieved.\par
In our case most of the running period was done with M$_{trigger}^{min}$=4. Correct sampling check is done by applying, off-line, 
the exclusive trigger condition to minimum bias trigger (M$_{trigger}^{min}$=1) recorded events and by comparing the selected 
impact parameter evaluator distribution to the original one. 
The sampling correctness is thus valued over the whole impact parameter range.\par
%%%%%%%%%%%%%
\begin{figure}[htbp]
\centering
\resizebox{0.45\textwidth}{!}{%
   \includegraphics{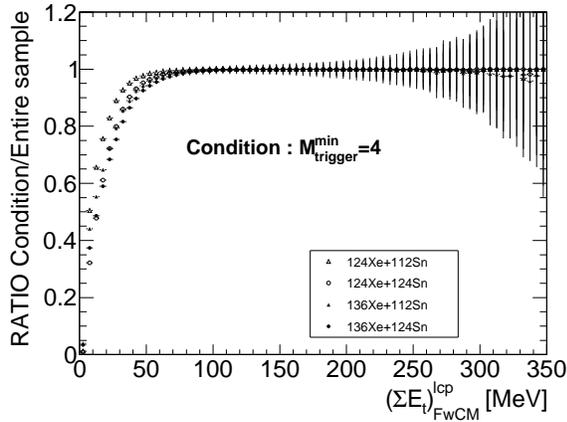}
   }  
   \caption{Ratio between $(\Sigma E_{t})_{FwCM}^{lcp}$ distributions of M$_{trigger}^{min}$=4 (off-line condition) and M$_{trigger}^{min}$=1 for the four studied reactions. The minimum bias samples were used.}
   \label{RepresentativiteM4}
   % code figure FiguresArbres_XeSnRunsSingles.C pour root : 2.1
\end{figure}
%%%%%%%%%%%%%%
The ratio of the two distributions is presented in figure \ref{RepresentativiteM4} for the four studied systems. It is seen
that for M$_{trigger}^{min}$=4 condition, the level of eliminated events is kept below 3\% for 
$(\Sigma E_{t})_{FwCM}^{lcp}$ greater than 70 MeV for the four systems. Below this value the M$_{trigger}^{min}$=4 sampling depends on impact parameter evaluator and is also system dependent. This implies studying very 
peripheral (b/b$_{max}$ greater than about 0.7) events is not possible with M$_{trigger}^{min}$=4 running condition.
\par 
Most of the data taking was performed with M$_{trigger}^{min}$=4 condition and thus for statistics purpose it may be convenient 
to mix inclusive and exclusive samples. A correct sampling ensemble is realized  by mixing  M$_{trigger}^{min}$=1 
sample for $(\Sigma E_{t})_{FwCM}^{lcp}$ lower than 70 MeV and M$_{trigger}^{min}$=4 sample for 
$(\Sigma E_{t})_{FwCM}^{lcp}$ greater than 70 MeV. 
In the following this mixed sample will be used when necessary.\par
%%%%%%%%%%%%%%%%%%%%%%%%%%%%%%%%%%%%%%%%%%%%%%%%%%%%%%%%
%%%%%%%%%%%%                                                                                                    SECTION
%%%%%%%%%%%%%%%%%%%%%%%%%%%%%%%%%%%%%%%%%%%%%%%%%%%%%%%%
\section{Light charged particle production}
\begin{figure}[htbp]
\centering
\resizebox{0.45\textwidth}{!}{%
   \includegraphics{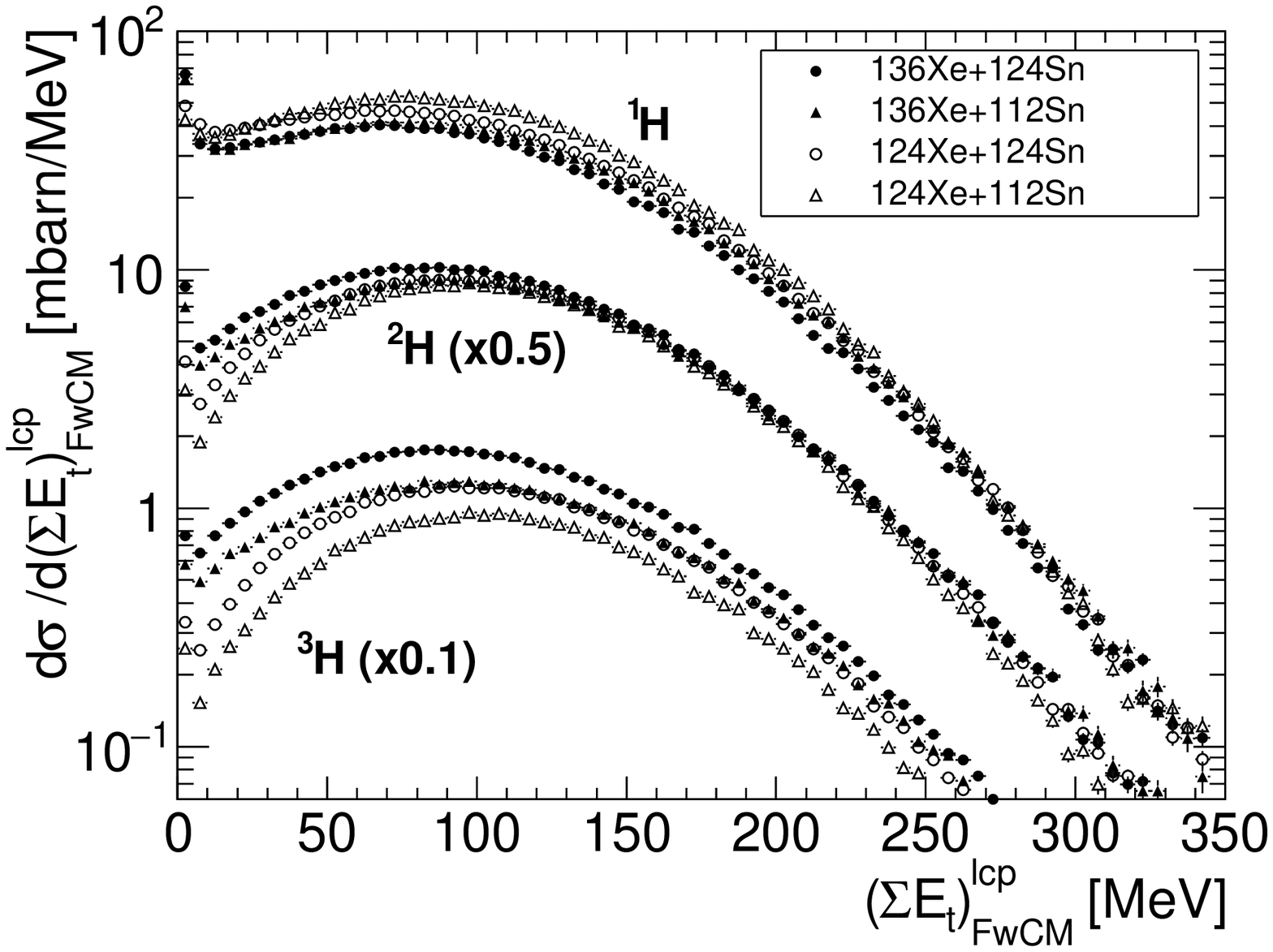}
   % code figure FiguresArbres_XeSnRunsSingles.C pour root : 3.8
   }
\resizebox{0.45\textwidth}{!}{%
   \includegraphics{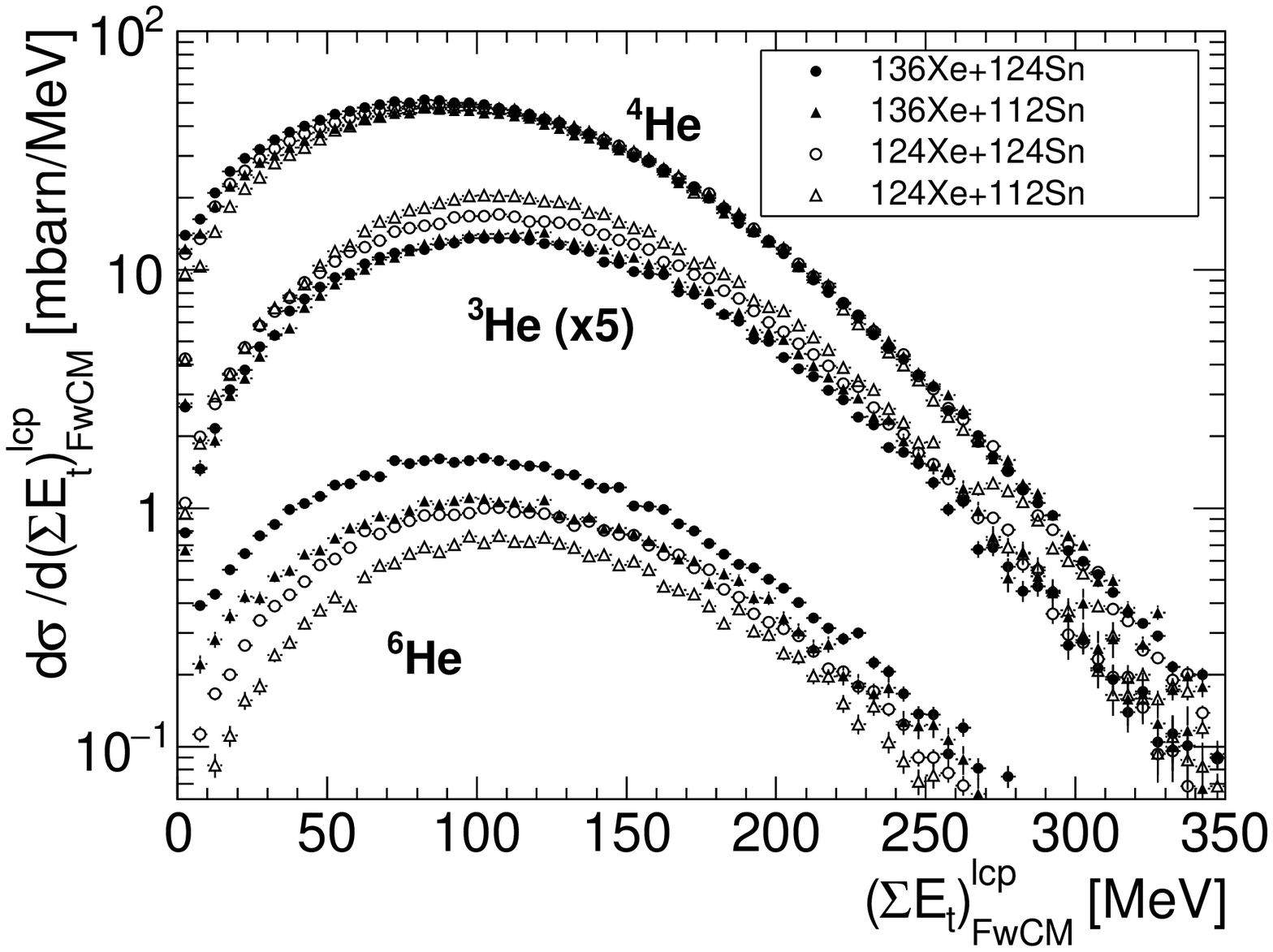}
   % code figure FiguresArbres_XeSnRunsSingles.C pour root : 3.9
   }    
   \caption{Forward center of mass light charged isotope productions for each studied reaction as a function of the 
impact parameter evaluator (minimum bias samples).}
   \label{LcpProductionMbarn}
\end{figure}

The forward center of mass lcp production is displayed in figure \ref{LcpProductionMbarn} for each studied system as a function of the impact parameter evaluator. 
%%%%%%%%%%
%%%%%%%%%%
For very peripheral collisions lcp production is largely dominated by $^{1}$H production. $^{1}$H production 
decreases and is partly replaced by cluster emission for smaller impact parameters while copious lcp production is achieved 
for more central collisions ($(\Sigma E_{t})_{FwCM}^{lcp}$ about 100 MeV). This implies that global cross sections given in table \ref{TableCrossSectionlcp} are largely influenced by lcp production around 0.5 reduced impact parameter. The figure indicates three types of behavior against N/Z:
(i) $^{3}$He production is projectile dependent for almost all impact parameters, 
(ii) symmetric lcp ($^{2}$H and $^{4}$He) production evolves from projectile dependence to system independence from peripheral to central reactions,
(iii) n-rich lcp ($^{3}$H and $^{6}$He) and $^{1}$H productions evolves from projectile dependence to combined system dependence from peripheral to central reactions.\par
Cross-section values reflect the production probabilities folded by the reaction cross-section. Therefore those values
cannot be used directly to study chemical composition of the four exit channel reactions.
Nevertheless from figure \ref{EtlcpAvecZoom_bsurbmax}-left, it was noticed an identical $^{124}$Xe+$^{124}$Sn and  $^{136}$Xe+$^{112}$Sn system reaction cross-sections for $(\Sigma E_{t})_{FwCM}^{lcp}>$ 60 MeV, whereas all studied system reaction cross sections are the same for $(\Sigma E_{t})_{FwCM}^{lcp}>$ 150 MeV. 
This means that from a simple direct measurement, figure \ref{LcpProductionMbarn}, it is possible to extract informations concerning N/Z equilibration selecting forward part center of mass emitted lcp. 
\par
%%%%%%%%%%%%%
\begin{figure}[htbp]
\centering
\resizebox{0.45\textwidth}{!}{%
 \includegraphics{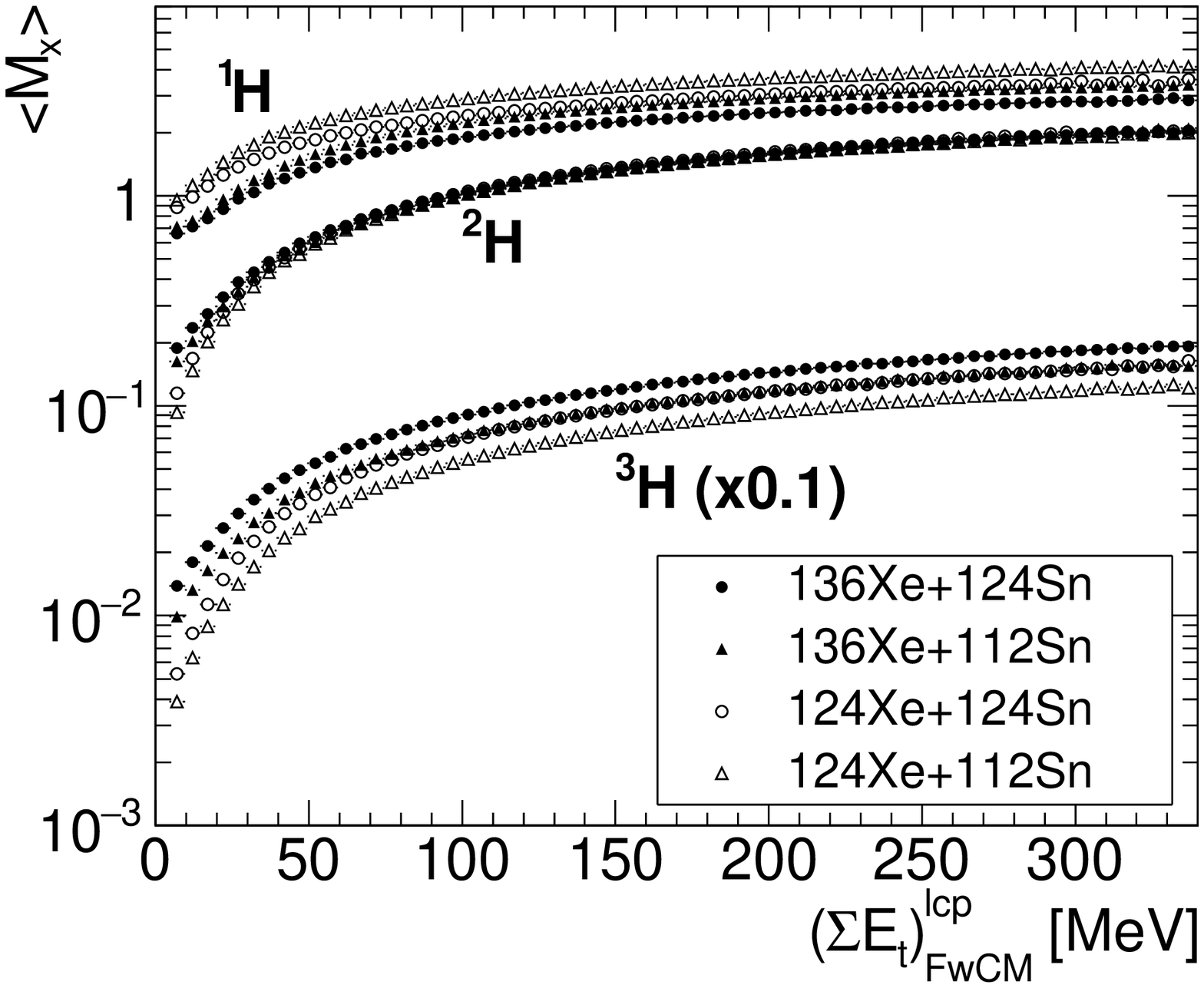}
   % code figure FiguresArbres_XeSnMelangeSingleMsup.C pour root : 801.1
  }
\resizebox{0.45\textwidth}{!}{%
 \includegraphics{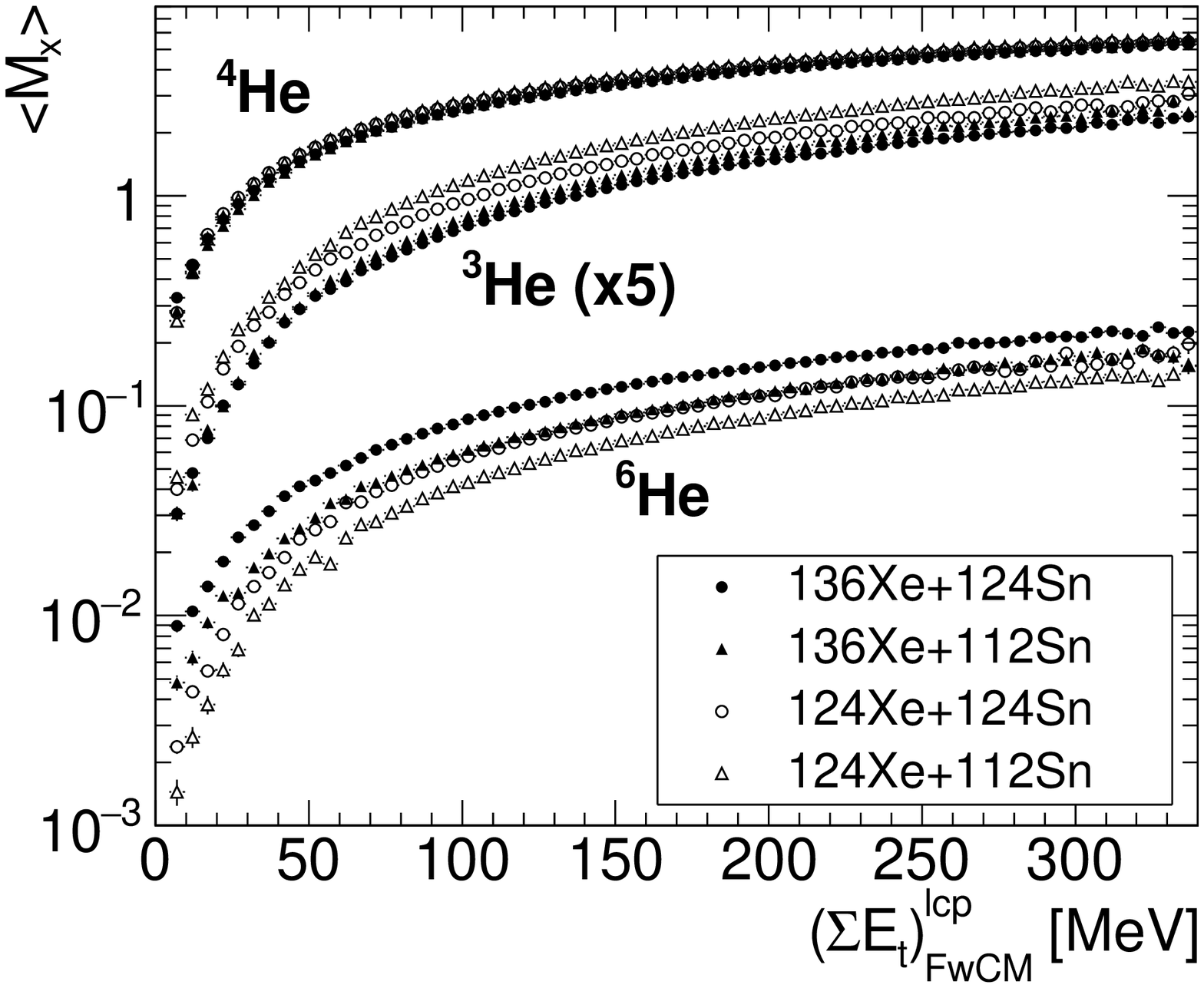}
   % code figure FiguresArbres_XeSnMelangeSingleMsup.C pour root : 801.2
  }    
   \caption{Forward center of mass light charged isotope mean multiplicities for each studied reaction as a function of the 
impact parameter evaluator (mixed samples).}
   \label{MlcpvsEtlcp}
\end{figure}
%%%%%%%%%%%%%%
The lcp yields were presented in 
figure \ref{LcpProductionMbarn}. The production probabilities, 
thus unfold by the reaction cross-section, are presented in figure \ref{MlcpvsEtlcp} 
supplied as mean multiplicities relative to the impact parameter evaluator. 
It is seen that all mean multiplicities increase with decreasing impact parameter. 
Decreasing the impact parameter, the whole system is more and more excited and particle multiplicities increase. Because the impact parameter evaluator and lcp multiplicities are self-correlated, multiplicity increasing never ceases in figure \ref{MlcpvsEtlcp}.
By comparing the values between the four systems it is possible to extract general evolutions. For very peripheral collisions, data are grouped in two categories (black points and white points, $^{136}$Xe and $^{124}$Xe projectiles respectively). In that case, the multiplicity evolution depends on the nature of the projectile. 
Decreasing the impact parameter, particle production deviates from this first order target independent behavior towards a dependence on the combined (projectile+target) system N/Z. 
This is evidenced by almost identical production, for central collisions, of most of the isotopes for $^{124}$Xe+$^{124}$Sn and $^{136}$Xe+$^{112}$Sn systems. This evolution is visible because only forward c.m. lcp production is shown.
For $^{2}$H and $^{4}$He multiplicities evolve towards almost identical mean values for the four studied systems. $^{3}$He multiplicity behavior is different since it
presents a trend linked to memory of projectile N/Z for all impact parameters.
\par
\begin{figure}[htbp]
\centering
\resizebox{0.45\textwidth}{!}{%
   \includegraphics{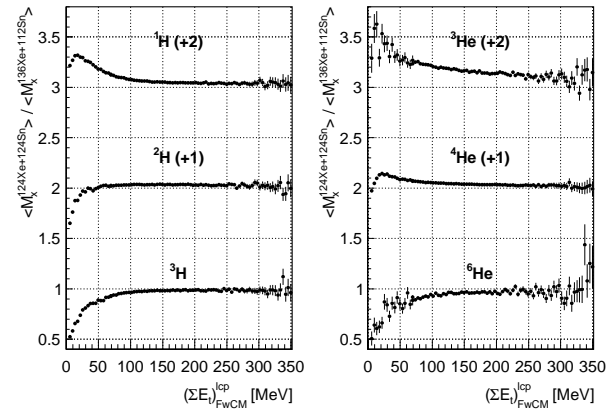}
   % code figure FiguresArbres_XeSnMelangeSingleMsup.C pour root : 901.31 (option 6)
   }  
   \caption{Ratios of mean light charged isotope multiplicities between $^{124}$Xe+$^{124}$Sn and $^{136}$Xe+$^{112}$Sn systems. In order to present the data in two pictures, a constant value (1 or 2) is added to certain ratios (mixed samples).}
   \label{MxRapportsSystemesvsEtlcp}
\end{figure}

Figure \ref{MxRapportsSystemesvsEtlcp} displays the ratios of mean multiplicities between the two identical global N/Z systems as a function of the impact parameter evaluator.   
%%%%%%%%%%%%%
%%%%%%%%%%%%%%
%The  $^{124}$Xe+$^{124}$Sn and $^{136}$Xe+$^{112}$Sn systems are used in order to compared particle %production for two systems whose total N/Z is identical. 
The ratios present the evolution from peripheral to central events with the n-rich projectile system in the denominator thus n-rich particle ratios evolve from below unity towards unity values. This evolution is inverted for n-poor particle ratios while symmetric particle ratios do not present an identical evolution as a function of centrality. Except for $^{3}$He, as previously stated, all ratios are close to unity for $(\Sigma E_{t})_{FwCM}^{lcp}$ greater than about 130 MeV.
\par
For very peripheral collisions the $^{1}$H and $^{4}$He ratios present a striking behavior since they first increase and then decrease with the impact parameter evaluator. In this study, peripheral reactions leading to only neutron production are excluded and therefore the mean lcp multiplicities of figure \ref{MlcpvsEtlcp} are overestimated if that figure is interpreted as mean multiplicities versus impact parameter. Obviously this overestimation is depending on the neutron richness of the projectile and not only  $^{1}$H and $^{4}$He multiplicities are affected. The figure \ref{MlcpvsEtlcp} is correct, only its representation in term of impact parameter is not pertinent for $(\Sigma E_{t})_{FwCM}^{lcp}$ below about 30 MeV. Above this value, it is expected that all reaction processes produce at least one charged particle. 
%%%%%%%%%%%%%%%%%%%%%%%%%%%%%%%%%%%%%%%%%%%%%%%%%%%%%%%%
%%%%%%%%%%%%                                                                                                    SECTION
%%%%%%%%%%%%%%%%%%%%%%%%%%%%%%%%%%%%%%%%%%%%%%%%%%%%%%%%
\section{Cluster abundance ratios}
In the previous section it was indicated a mean multicity overestimation problem for very peripheral reactions if we think in terms of impact parameter dependence.
One way round this is to compare cluster mean multiplicities relative to proton mean multiplicity (hereafter called cluster abundance ratios \cite{GutbrodRepProgPhys52}). Doing so, for each system the multiplicity overestimation seen for very peripheral collisions is canceled and it is then possible to compare cluster abundance ratios whatever the impact parameter is.
\par 
The use of cluster abundance ratios allows also to study chemical equilibration process because trivial size dependences are removed \cite{ReisdorfNPA848}. In case of chemical equilibrium, one expects that both  $^{124}$Xe+$^{124}$Sn and  $^{136}$Xe+$^{112}$Sn systems lead to same cluster concentration for a given species so same abundance ratio. $^{3}$He characteristics will be studied in the next section because of their previously noticed peculiar behavior.
\par
\begin{figure}[htbp]
\centering
\resizebox{0.45\textwidth}{!}{%
   \includegraphics{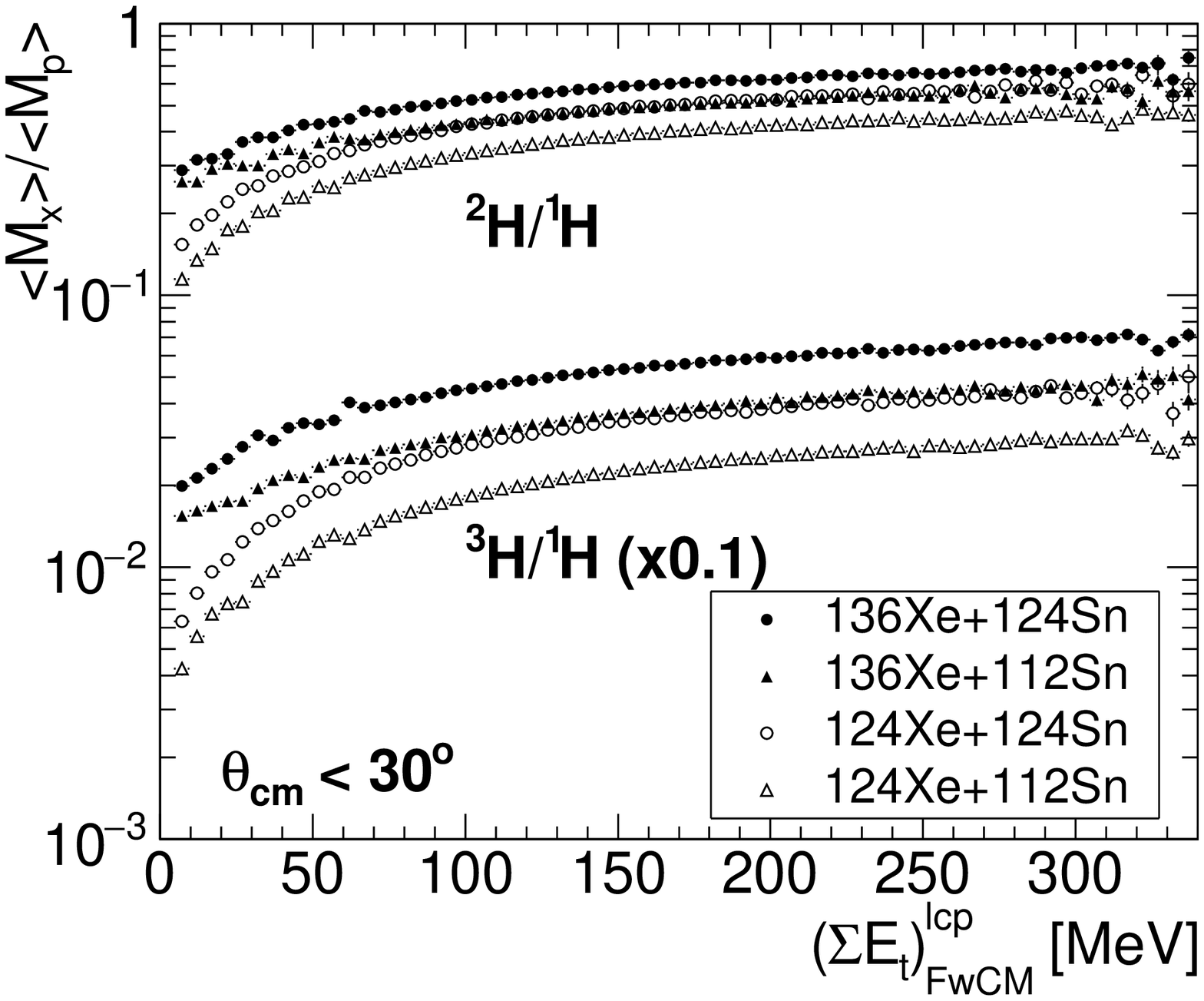}
   % code figure FiguresArbres_XeSnMelangeSingleMsup.C pour root : 900 (puis choix option 0-30 etc)
   }  
\resizebox{0.45\textwidth}{!}{%
   \includegraphics{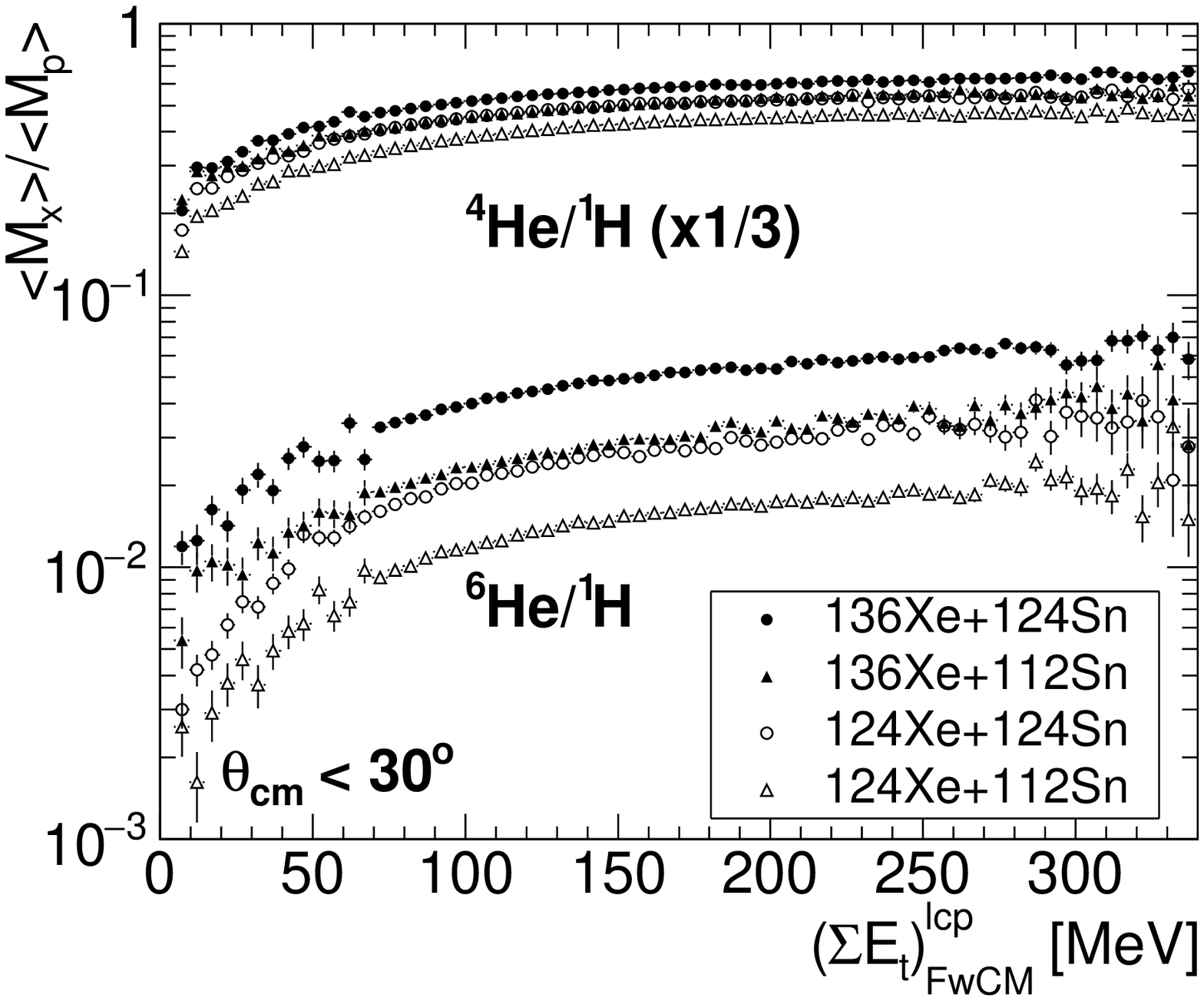}
   % code figure FiguresArbres_XeSnMelangeSingleMsup.C pour root : 900 (puis choix option 0-30 etc)
   } 
    \caption{Forward center of mass cluster abundance ratios for each studied reaction as a function of the impact parameter evaluator (mixed samples). 
Particles emitted forward (0$^{\rm{o}}$-30$^{\rm{o}}$).}
\label{MxsurMp0030vsEtlcp}
\end{figure}
\begin{figure}[htbp]
\centering   
\resizebox{0.45\textwidth}{!}{%
   \includegraphics{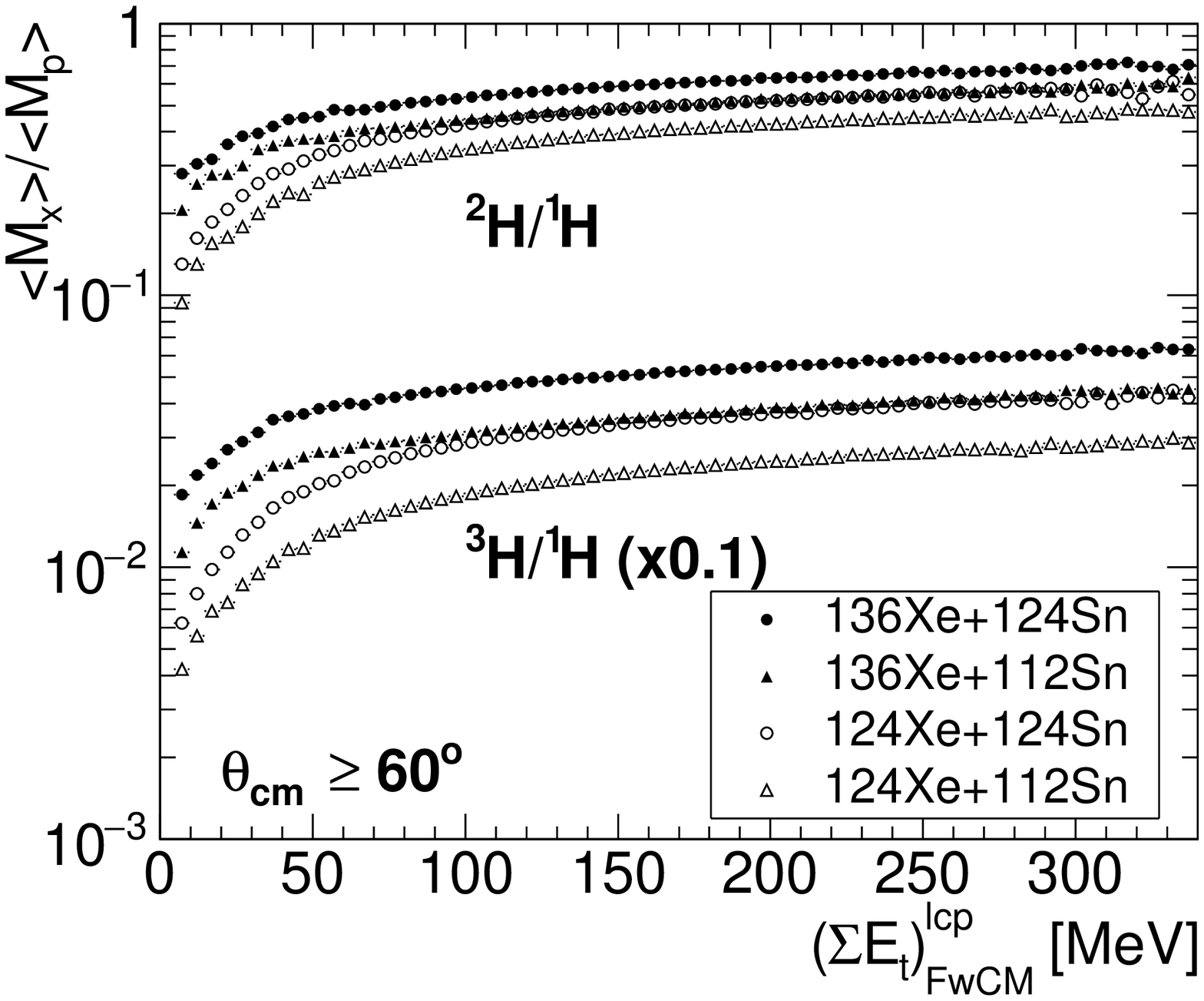}
   % code figure FiguresArbres_XeSnMelangeSingleMsup.C pour root : 900 (puis choix option 0-30 etc)
   } 
\resizebox{0.45\textwidth}{!}{%
   \includegraphics{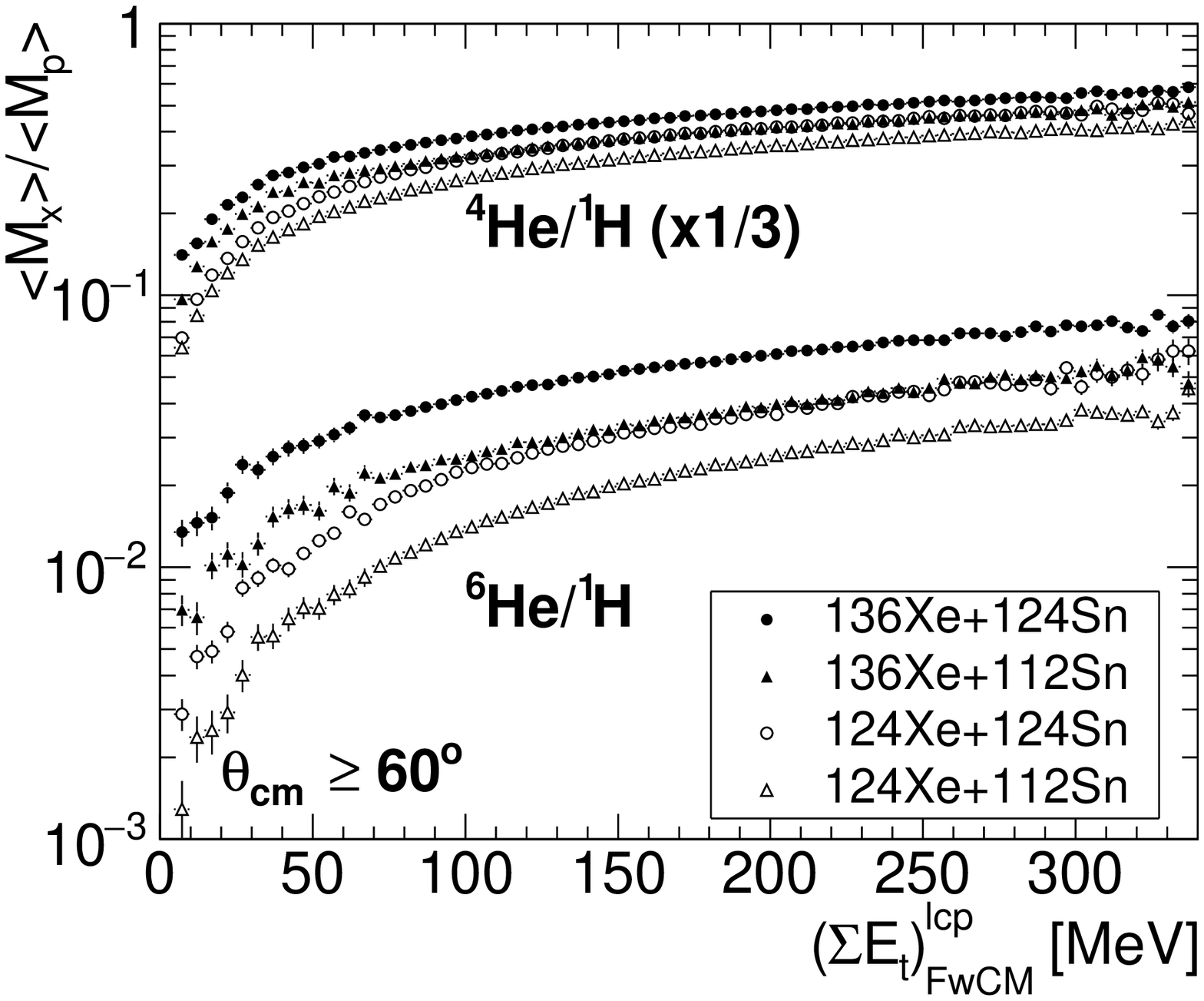}
   % code figure FiguresArbres_XeSnMelangeSingleMsup.C pour root : 900 (puis choix option sup. 60 etc)
   }  
   \caption{Forward center of mass cluster abundance ratios for each studied reaction as a function of the impact parameter evaluator (mixed samples). 
Particles emitted perpendicular to the beam direction (60$^{\rm{o}}$-90$^{\rm{o}}$).}
\label{MxsurMp6090vsEtlcp}
\end{figure}

Cluster abundance ratios are presented in figures \ref{MxsurMp0030vsEtlcp} and \ref{MxsurMp6090vsEtlcp}. Two center of mass lcp polar angular ranges are selected in order to approximately select projectile-like de-excitation (0$^{\rm{o}}$-30$^{\rm{o}}$) and mid-rapidity (60$^{\rm{o}}$-90$^{\rm{o}}$) populations. The ratios are calculated using the total number of clusters and proton detected in the given angular range and in the given impact parameter evaluator bin. 
%%%%%%%%%%%%%
%%%%%%%%%%%%%%
\par
From the figures we conclude the following:
\begin{itemize}
\item Mean cluster abundance ratios are increasing with centrality. This implies an increase of composite particle production caused by an increase of excitation energy and nucleon-nucleon collisions.
\item For $(\Sigma E_{t})_{FwCM}^{lcp}$ greater than about 150 MeV, $^{124}$Xe+$^{124}$Sn and  $^{136}$Xe+$^{112}$Sn system mean abundance ratios are almost the same for a given cluster and a given angular range. This global N/Z system dependence implies that chemical equilibrium is almost achieved for central collisions (reduced impact parameters lower 
about 0.3).
\item Cluster abundance ratio evolution against impact parameter evaluator reflects the dynamical process which occurs during the collision. For projectile-like de-excitation region, the evolution starts from almost N/Z projectile dependence to N/Z total system dependence. For mid-rapidity region the values are also projectile N/Z dependent for very peripheral reactions
%, but slightly loosely as compared to projectile-like de-excitation region, 
whereas they also reach a N/Z total system dependence for central collisions. This evolution reflects the drift/diffusion isospin phenomena and the mid-rapidity population behavior cannot be described by a pure participant/spectator scenario \cite{WestfallPRL37}.
\end{itemize}
The two reactions
$^{124}$Xe+$^{124}$Sn and  $^{136}$Xe+$^{112}$Sn 
leading to the same projectile+target combined system were chosen to study the path towards chemical equilibrium. Comparing mean cluster abundance ratios we did not measure exactly the same values for the two systems. The original idea was neglecting pre-equilibrium particle emission which could be different for the two systems and thus explain the slight measured differences (few \%). 
We will demonstrate this point in the next paragraph using $^{3}$He production. Nevertheless the fact remains that measured cluster abundance ratio values between the two systems are so close, comparing to $^{124}$Xe+$^{112}$Sn and  $^{136}$Xe+$^{124}$Sn systems, that the
assumption of chemical equilibrium achievement is justified since abundance ratios are largely global (projectile+target) N/Z dependent. 
%%%%%%%%%%%%%%%%%%%%%%%%%%%%%%%%%%%%%%%%%%%%%%%%%%%%%%%%
%%%%%%%%%%%%                                                                                                    SECTION
%%%%%%%%%%%%%%%%%%%%%%%%%%%%%%%%%%%%%%%%%%%%%%%%%%%%%%%%
\section{The helion case}
The peculiar characteristic of $^{3}$He, as compared to other lcp, produced in collisions between heavy targets with proton, light
or heavy projectiles has been pointed out in \cite{ReisdorfNPA848} \cite{PoskanzerPRC3}  \cite{XiPRC58} \cite{MariePLB391} \cite{NeubertEPJA7}  and was previously indicated in this article.
%\ref{LcpProductionMbarn}, \ref{MlcpvsEtlcp} and \ref{MxRapportsSystemesvsEtlcp}. 
%%%%%%%%%%%%%
\begin{figure}[htbp]
\centering
\resizebox{0.45\textwidth}{!}{%
   \includegraphics{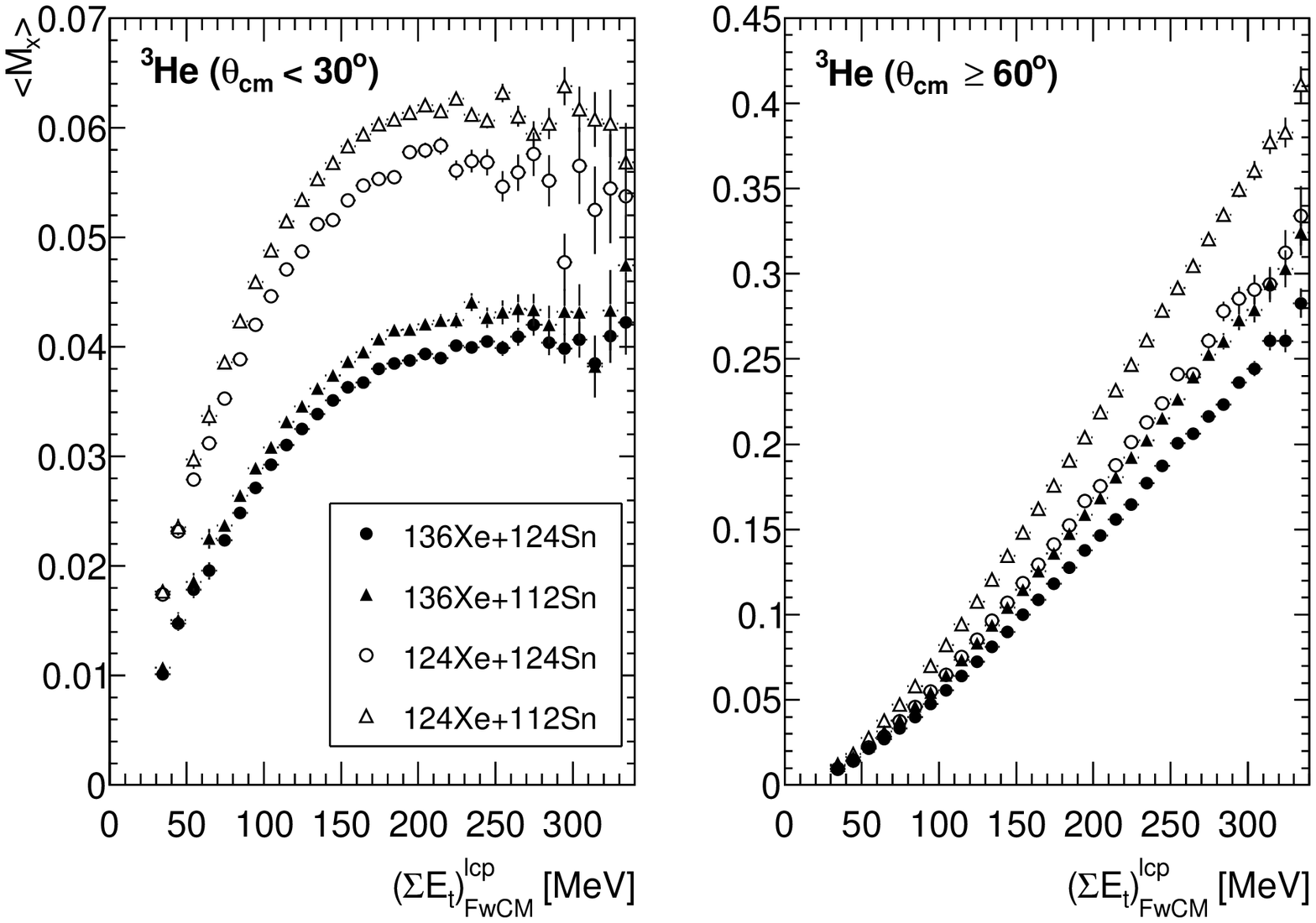}
   % code figure FiguresArbres_XeSnMelangeSingleMsup.C pour root : 909.2
   }  
%%%%%%%%%%%%%%
%%%%%%%%%%%%%
\centering
\resizebox{0.45\textwidth}{!}{%
   \includegraphics{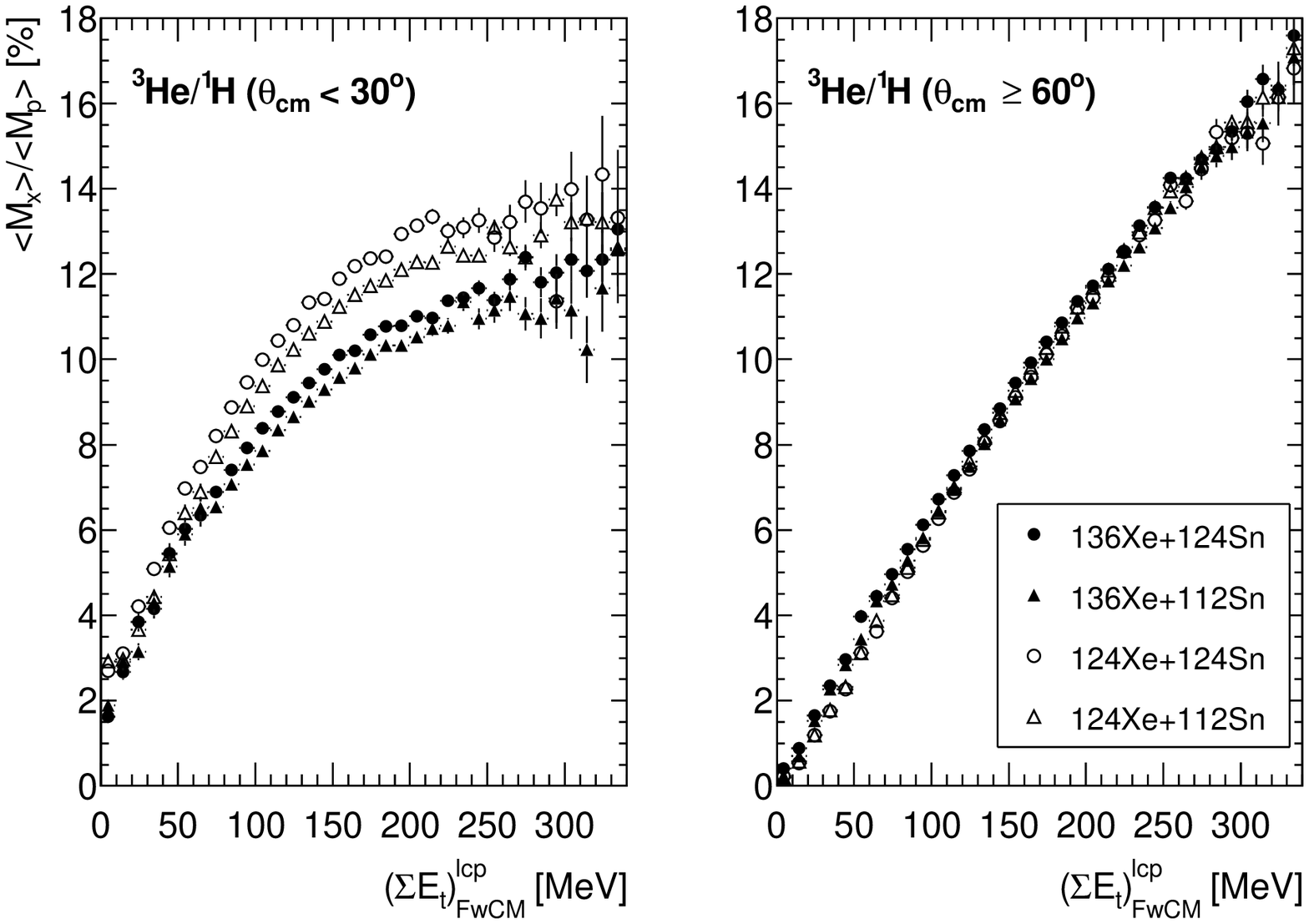}
   % code figure FiguresArbres_XeSnMelangeSingleMsup.C pour root : 909.1
   }  
   \caption{Forward center of mass $^{3}$He mean multiplicities (top) and abundance ratios (bottom) for each studied 
reaction as a function of the impact parameter evaluator (mixed samples). Abundance ratio values are in \%.
Left: particles emitted forward (0$^{\rm{o}}$-30$^{\rm{o}}$). Right: particles emitted perpendicular to the beam direction (60$^{\rm{o}}$-90$^{\rm{o}}$).}
   \label{M3He3060vsEtlcp}
\end{figure}
%%%%%%%%%%%%%%
\par
\par
Figure  \ref{M3He3060vsEtlcp} shows studied system mean helion multiplicities and abundance ratios for the two center of mass lcp polar angular ranges (projectile-like de-excitation and mid-rapidity populations) as a function of impact parameter evaluator. 
The figure shows that helion production is very different as compared to other lcp presented in figures \ref{MxsurMp0030vsEtlcp} and \ref{MxsurMp6090vsEtlcp}. First, for all impact parameters the behavior between 0$^{\rm{o}}$-30$^{\rm{o}}$ and 60$^{\rm{o}}$-90$^{\rm{o}}$ populations is not the same. Secondly, when comparing the different system productions it is also observed that:
\begin{itemize}
\item The mean multiplicities remain largely projectile dependent for the projectile-like population while it is total system (projectile+target) dependent for the mid-rapidity population. This is true whatever the impact parameter is. In both cases n-poor system favors helion production.
\item Looking at $^{3}$He abundance ratios one observes that (i) projectile-like population remains largely projectile dependent for the whole impact parameter range, (ii) mid-rapidity population is independent of the reaction system for all impact parameters except to a certain extent for very peripheral collisions.
\end{itemize}
Putting together multiplicity and abundance ratio results, it is then observed that helion production conserves a footprint of initial projectile N/Z for the  0$^{\rm{o}}$-30$^{\rm{o}}$ domain while it depends on the size (not N/Z) of the overlapping region between the projectile and the target for the 60$^{\rm{o}}$-90$^{\rm{o}}$ domain.
\par
Isospin diffusion and drift are driving the chemical equilibrium process.
The latter phenomenon causes a neutron enrichment of the mid-rapidity zone (\cite{LarochellePRC62},  \cite{BarliniPRC87} and references therein) while the isospin diffusion tends to N/Z equilibrium between projectile and target collision partners. The observed results imply an average helion production prior chemical equilibrium achievement. For the case of mid-rapidity region, the neutron enrichment which occurs for all studied reactions disadvantages helion production when it becomes efficient. 
Keeping in mind that mean values are studied, these observations do not imply that all helion are produced before drift and diffusion mechanisms become fully effective, they rather imply that helion production is strongly reduced when those mechanisms acts.
\par
The mean pre-equilibrium character of $^{3}$He and its production difference between the studied systems for the 0$^{\rm{o}}$-30$^{\rm{o}}$ domain would appear to explain the small measured abundance ratio differences between $^{124}$Xe+$^{124}$Sn and  $^{136}$Xe+$^{112}$Sn systems. 
Other lcp are certainly partly concerned by pre-equilibrium production which gives rise to transparency effect \cite{LopezPRC14} but in average other lcp do not present this character.
%%%%%%%%%%%%%%%%%%%%%%%%%%%%%%%%%%%%%%%%%%%%%%%%%%%%%%%%
%%%%%%%%%%%%                                                                                                    SECTION
%%%%%%%%%%%%%%%%%%%%%%%%%%%%%%%%%%%%%%%%%%%%%%%%%%%%%%%%
\section{Mid-rapidity neutron enrichment and the $^{6}$He case}
\begin{figure}[htbp]
\centering
\resizebox{0.45\textwidth}{!}{%
   \includegraphics{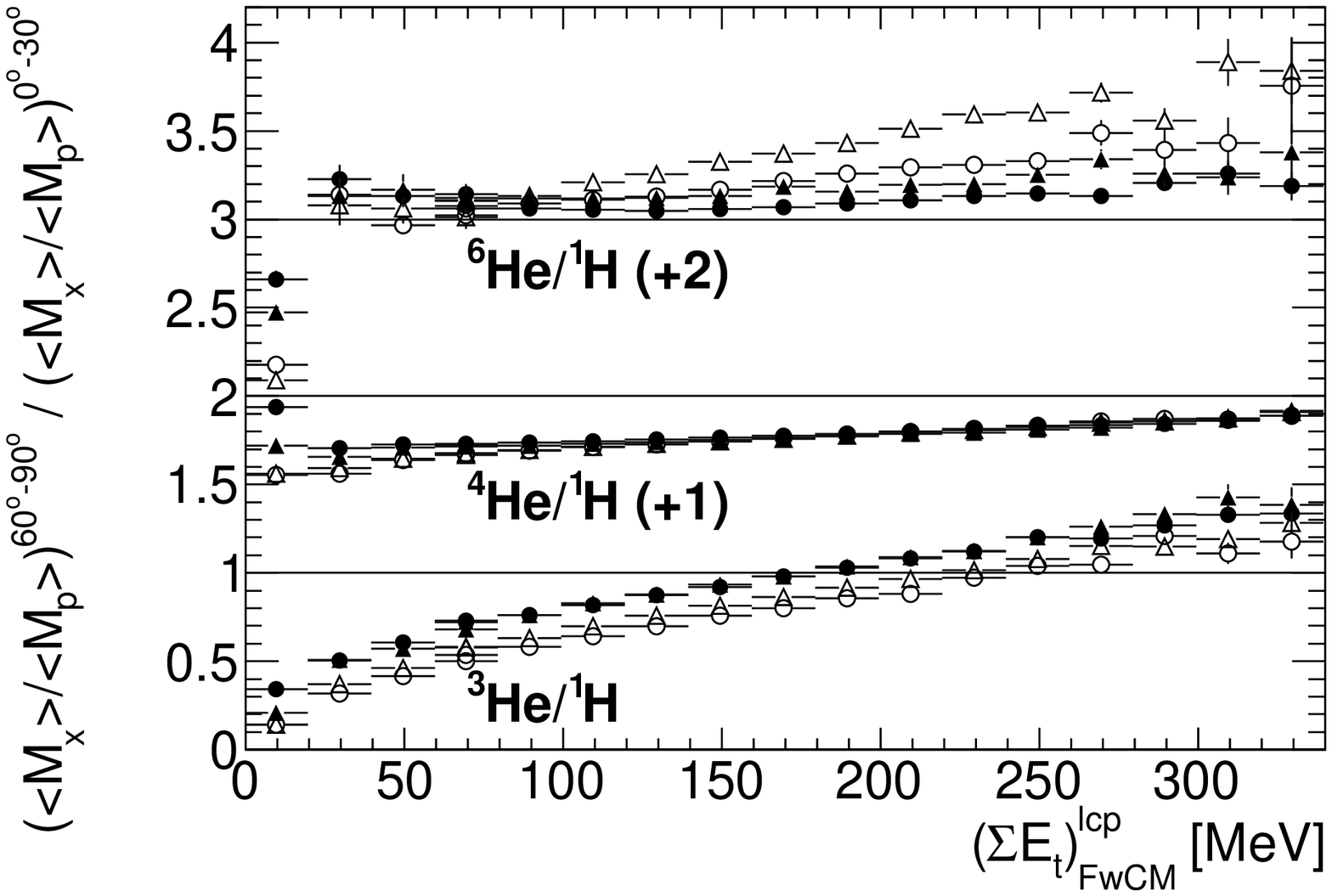}
   % code figure FiguresArbres_XeSnMelangeSingleMsup.C pour root : 900.1 (mettre des commentaires, deux passages pour faire deux figures)
   } 
\resizebox{0.45\textwidth}{!}{%
   \includegraphics{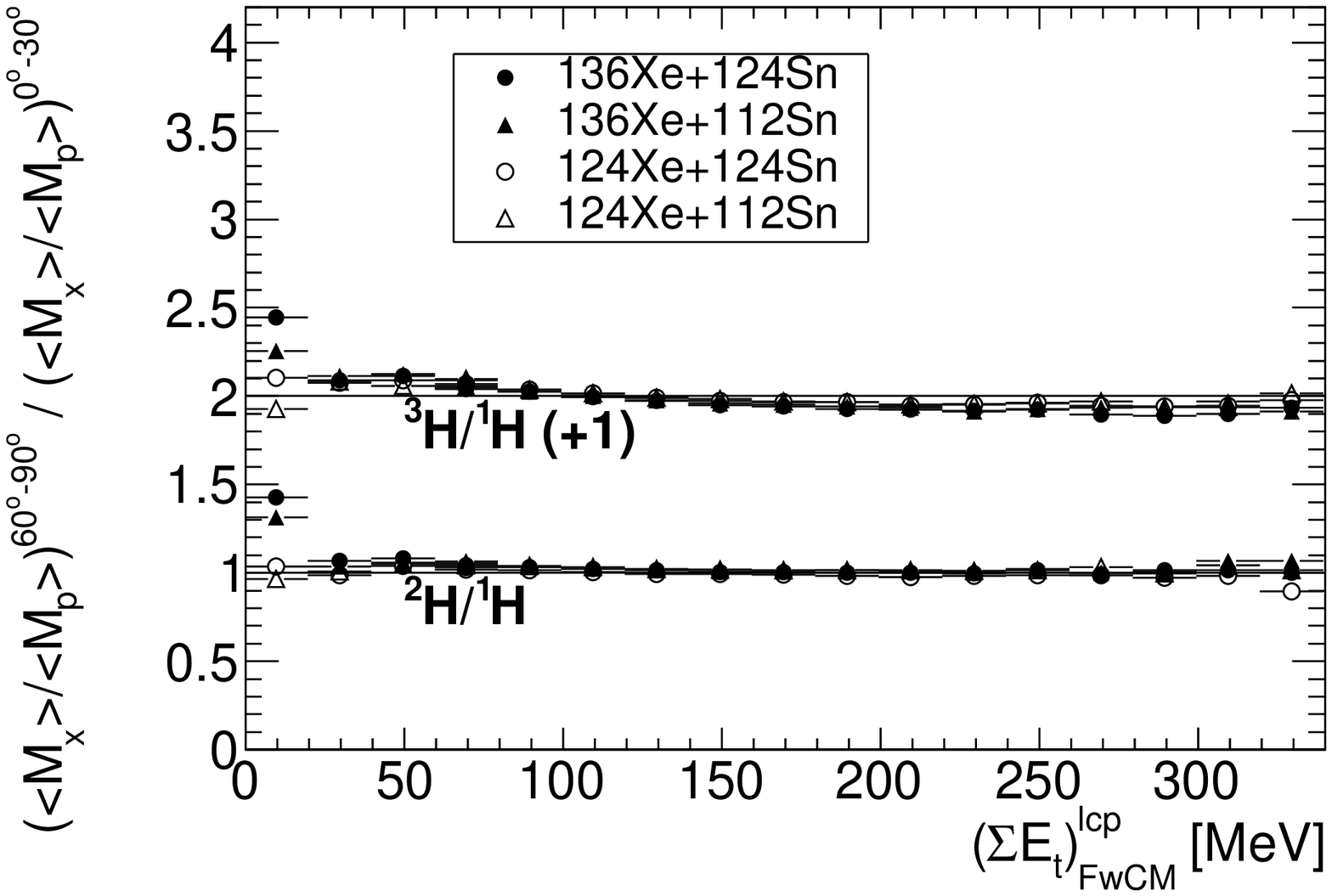}
   }    
   \caption{Forward center of mass $^{2}$H, $^{3}$H, $^{3}$He, $^{4}$He, $^{6}$He fractions for each studied 
reaction as a function of the impact parameter evaluator (mixed samples). Lcp fractions are simply 60$^{\rm{o}}$-90$^{\rm{o}}$ divided by 0$^{\rm{o}}$-30$^{\rm{o}}$ lcp abundance ratios. In order to present the data in two pictures, a constant value (1 or 2) is added to certain fractions.}
   \label{AbundanceRatio6090sur0030}
\end{figure}

Solid angle independence of cluster abundance ratios permits to compare directly the projectile-like and mid-rapidity mean values previously presented. The fraction between mean values of mid-rapidity and projectile-like angular regions against the impact parameter evaluator are presented in figure \ref{AbundanceRatio6090sur0030} for H and He isotopes.
\par
%%%%%%%%%%%%%
%%%%%%%%%%%%%%
\begin{itemize}
\item Concerning the system dependence, $^{3}$He fraction values reflect the 0$^{\rm{o}}$-30$^{\rm{o}}$ abundance ratio behavior since $^{3}$He abundance ratios are system independent for the 60$^{\rm{o}}$-90$^{\rm{o}}$ populations. It is not the case for the other particles. 
\item $^{2}$H,  $^{3}$H and $^{4}$He fractions are system independent for central collisions.
Except for peripheral reactions, the $^{2}$H fractions for the two populations (0$^{\rm{o}}$-30$^{\rm{o}}$ and 60$^{\rm{o}}$-90$^{\rm{o}}$) are identical. To a certain extent this is also true for $^{3}$H fractions. For $^{4}$He fractions,
some differences are observed at low $(\Sigma E_{t})_{FwCM}^{lcp}$ values and the increasing behavior with impact parameter evaluator may reflect angular momentum effects since $^{4}$He can remove appreciable
angular momentum from the excited projectile-like.
\item $^{6}$He fraction values are increasing with impact parameter evaluator as $^{4}$He but $^{6}$He fraction values are always larger than unity except for very peripheral collisions. This means that mid-rapidity region favors very neutron rich cluster production whatever the impact parameter is. Concerning the system dependence, the figure indicates that the more n-poor system is, the greater the fraction value. 
\end{itemize}
Besides $^{3}$He pre-equilibrium nature, those observations imply two different mean particle production modes for all impact parameters: projectile-like de-excitation and mid-rapidity sources whose N/Z are different for a given system and a given impact parameter. 
$^{2}$H and $^{3}$H fraction values may imply that, in average, projectile-like de-excitation process is largely dominating the production for reduced impact parameters below around 0.6.
The drift phenomenon may explain the system dependence of $^{6}$He fractions: the n-enrichment of mid-rapidity source dry out the projectile-like from sufficient neutron concentration to produce very rich neutron clusters, the more n-poor the system is, the more dramatic is the effect. For $^{124}$Xe+$^{112}$Sn reaction this cannot be counterbalanced by the diffusion effect since the projectile and target N/Z are similar.  
%%%%%%%%%%%%%%%%%%%%%%%%%%%%%%%%%%%%%%%%%%%%%%%%%%%%%%%%
%%%%%%%%%%%%                                                                                                    SECTION
%%%%%%%%%%%%%%%%%%%%%%%%%%%%%%%%%%%%%%%%%%%%%%%%%%%%%%%%
\section{Conclusion}
We have presented INDRA multi-detector data acquired through 32 MeV/nucleon $^{136,124}$Xe+$^{124,112}$Sn reactions.
The study was restricted to forward part of the center of mass emitted light charged particles because for those particles the excellent detection performance allows to carry out inclusive analysis. Two studied reactions
($^{124}$Xe+$^{124}$Sn and  $^{136}$Xe+$^{112}$Sn) were chosen to study the degree of chemical equilibrium, i.e the
N/Z balance between the projectile and the target. This balance was studied for all type of collisions by estimating the 
impact parameter. Shown results concern average behaviors and no restricted selection was performed. 
\par
Light charged particle productions, multiplicities and abundance ratios dependence against impact parameter indicate
that a high degree of chemical equilibrium is achieved in central collisions. This conclusion was established by measuring
almost identical mean characteristics for the two $^{124}$Xe+$^{124}$Sn and $^{136}$Xe+$^{112}$Sn systems 
which are different when comparing with $^{124}$Xe+$^{112}$Sn and $^{136}$Xe+$^{124}$Sn systems: mean values
are projectile+target N/Z dependent.
\par
This high degree of equilibration is of the order of few \% difference for $^{124}$Xe+$^{124}$Sn and 
$^{136}$Xe+$^{112}$Sn systems. This slight difference could be explained by pre-equilibrium particle emission whose
intensity may differ for the two reactions. This point has been demonstrated using $^{3}$He mean characteristics which strongly differ from other lcp behaviors. It has been shown that helion production takes place before chemical equilibrium
achievement.
\par
The achieved N/Z balance between the projectile and target does not imply a pure two-body mechanism. Indeed a mid-rapidity
source of lcp production does exist and its N/Z is different as compared to the projectile-like one: it is n-enriched. This point has been touched using $^{6}$He production which is favored by the drift phenomenon.
\par
Our results differ from those of \cite{KeksisPRC81} and \cite{SunPRC82}. We recall that here raw data are used while the cited
results are extracted using reconstructing methods or obtained through restricted rapidity region of the phase space.
\par
With the advent of the wide range mass and charge identification FAZIA detector \cite{BougaultEPJA50}, this study could be 
extended to higher elements.
\par
Finally we think that $^{6}$He and $^{3}$He productions may be the key for comparing data to transport model results concerning the knowledge of the equation of state and its isospin dependence.
%%%%%%%%%%%%
%%%%%%%%%%%%%
%%%%%%%%%%%%%%
%%%%%%%%%%%%%%%%%%%%%%%%%%%%%%%%%%%%%%%%%%%%%%%%%%%%%%%%
%%%%%%%%%%%%                                                                                                    SECTION
%%%%%%%%%%%%%%%%%%%%%%%%%%%%%%%%%%%%%%%%%%%%%%%%%%%%%%%%
%%%%%%%%%%%%%%%%%%%%%%%%  REFERENCES %%%%%%%%%%%%%%%%%%%%
%\bibliographystyle{plain}
\bibliography{BougaultPRC2017Bib}
% End of the paper
\end{document}